\documentclass[12pt]{article}
\usepackage{amsmath}
\usepackage{epsfig,graphicx}
\usepackage{graphicx}
\usepackage{tikz}
\def\be{\begin{equation}}
\def\ee{\end{equation}}
\def\bea{\begin{eqnarray}}
\def\eea{\end{eqnarray}}

\def\rme{\mathrm{e}}
\def\rmi{\mathrm{i}}
\def\scq{{\textsc q}}

\begin{document}

\begin{center}
{\large \bf The algebraic area of closed lattice random walks}\vskip 0.5cm

{\large \bf St\'ephane Ouvry}\footnote{stephane.ouvry@u-psud.fr}\\[0.1cm]
{\large \bf Shuang Wu}\footnote{shuang.wu@u-psud.fr}\\[0.1cm]

Laboratoire de Physique Th\'eorique et Mod\`eles Statistiques,
CNRS-Universit\'e Paris Sud, Universit\'e Paris-Saclay, Facult\'e des Sciences d'Orsay,
91405 Orsay, France

\today
\end{center}

\vskip 0.5cm
\centerline{\large \bf Abstract}
\vskip 0.2cm  
We propose   a formula for the enumeration  of  closed lattice random walks of length $n$  enclosing a given algebraic area. The information is contained in the Kreft coefficients which encode, in the commensurate case,  the  Hofstadter secular equation for a quantum particle hopping on a lattice coupled to a perpendicular magnetic field. 
 The  algebraic area  enumeration is possible because it is split in  $2^{n/2-1}$ pieces, each tractable in terms of explicit combinatorial expressions. 
%PACS numbers: 05.40.Fb, 05.30.Jp, 03.65.Aa \\

\section{Introduction}
The enumeration on a square lattice of closed  random walks  of length $n$, with $n$  then necessarily even,  { starting} from a given point and  enclosing a  given  algebraic area   seems as far as we can see still a current issue. We propose a formula for the enumeration  by splitting it in  $2^{n/2-1}$ pieces,  where $2^{n/2-1}$ is the number  of  partitions of $n/2$ where   partitions differing by the order of their parts are counted separately  ---e.g.,    $4=2+1+1$, $4=1+2+1$  and $4=1+1+2$ each count. One then refers to compositions rather than to partitions.  

The  observation  which allows for the  algebraic area enumeration  originates from the Hofstadter model for a quantum particle hopping on a square lattice and coupled to a perpendicular magnetic field.

The algebraic area is the area enclosed by a curve, weighted by its winding number: if the curve moves around a region in counterclockwise  direction, its area counts as positive, otherwise negative. Moreover, if the curve winds around more than once, the area is counted with multiplicity.  We focus on the algebraic area of  walks on a square lattice starting from a given point   and at each step moving right, left, up or down   with equal probability. 
\iffalse
\begin{figure}[htbp]
\begin{center}
\begin{tikzpicture}[scale=0.6]
\draw [dashed, fill=lightgray] (0,0)--(0,-2)--(3,-2)--(3,1)--(2,1)--(2,2)--(1,2)--(1,0)--(0,0);
\draw [very thick, fill=lightgray] (3,1)--(5,1)--(5,2)--(4,2)--(4,3)--(3,3)--(3,1);
\draw [very thick] (0,0)--(0,-2)--(3,-2)--(3,1)--(2,1)--(2,2)--(1,2);
\draw [very thick, fill=lightgray] (5,0)--(6,0)--(6,1)--(5,1)--(5,0);
\node[fill=black,circle,inner sep=2pt] () at (0,0) {};
\node at (1.5,-0.5) {$+$};
\node at (3.5,1.5) {$-$};
\node at (5.5,0.5) {$+$};
\end{tikzpicture}
\end{center}
\caption{A lattice walk with algebraic area $9-3+1$.}
\end{figure}
\fi

Suppose that a  walk has moved    $m_1$ steps right, $m_2$ steps left, $l_1$ steps up and $l_2$ steps down. If e.g. $m_1 \ge  m_2$ and $l_1 \ge l_2$, 
we add $l_1 - l_2$ steps down followed by $m_1 - m_2$ steps  left in order to close the walk and endow it  with an algebraic area. Let $C_{m_1,m_2,l_1,l_2}(A)$ be the number of such walks which enclose  a given algebraic area $A$. Finding the generating function for the $C_{m_1,m_2,l_1,l_2}(A)$'s
\be Z_{m_1,m_2,l_1,l_2}(\scq)=\sum_A C_{m_1,m_2,l_1,l_2}(A) \scq^A \nonumber\ee is quite challenging.
One restricts to closed lattice walks of length $n$ ($n$ is then necessarily even), i.e., walks with an equal number $m$ of steps right/left and an equal number $n/2-m$ of steps up/down,  $m \in \{0,1,\ldots,n/2\}$,   and    focuses on their algebraic area generating function  
\begin{equation} \label{Zn_def} Z_n(\scq)=\sum_{m=0}^{n/2}Z_{m, m, \frac{n}{2}-m, \frac{n}{2}-m}(\scq)=\sum_{A} C_{n}(A) \scq^A\end{equation}  where  $C_n(A)$ enumerates  closed  walks of length $n$  enclosing an algebraic area $A$  ($A$ is in between $-\lfloor n^2/16\rfloor$ and $\lfloor n^2/16\rfloor$ where $\lfloor\;\rfloor$  denotes the integer part;  obviously $C_{n}(A)=C_{n}(-A)$).

There is a connection  between the algebraic area distribution  of curves  and the quantum spectrum of a charged particle coupled to a perpendicular magnetic field. { This connection arises for continuous closed Brownian curves and their algebraic area distribution given by  L\'evy's law \cite{Levy}
\be P_t(A)={\pi\over 2t}{1\over \cosh(\pi A/t)^2}\;,\label{Levy}\ee  where $t$ is the time of the Brownian motion. One notes that $P_t(A)$ is nothing but the Fourier transform of the Landau partition function at inverse temperature $t$ of a quantum planar particle coupled to a perpendicular magnetic field. It is not a surprise that a magnetic field should  play a role  since it  indeed couples to the algebraic area spanned by paths in a  path integral formulation.} 
 In the lattice case  at hand,  the mapping  is on the quantum Hofstadter model \cite{Hofstadter} for a particle hopping on a two-dimensional lattice  coupled to  a  magnetic field with flux $\gamma$ per  lattice cell, in unit of the flux quantum.  $Z_n(e^{i\gamma})$  is mapped \cite{bellissard}  on the $n$-th moment ${\rm Tr}\:H_{\gamma}^{n}$ of the Hofstadter Hamiltonian $H_{\gamma}$
\be \label{thestuff} Z_{n}(e^{i\gamma})={\rm Tr}\:H_{\gamma}^{n}\ee
by virtue of which evaluating  $Z_n(e^{i\gamma})$  for  lattice walks gives an expression  for the  quantum trace ${\rm Tr}\:H_{\gamma}^{n}$, and vice versa.  { The coupling to a perpendicular magnetic field induces a non commuting lattice space  which in turns allows for  weighting discrete paths   by  their algebraic area \cite{bellissard}.}

${\rm Tr}\:H_{\gamma}^{n}$ can be  written \cite{nous} in terms of the Kreft coefficients \cite{Kreft} which    encode  the   Schrodinger equation for the Hofstadter model.
% \be \Phi_{m+1}+\Phi_{m-1}+2\cos(k_y+\gamma m)\Phi_{m}=E\Phi_{m}.\label{eq}\ee 
In the commensurate case with a rational   flux $
\gamma  = 2 \pi {p}/{q}$ 
 --$p$, $q$ coprime--  the Schrodinger equation reduces   to a $q\times q$ matrix  whose  determinant, more precisely its momentum independent part, can be expressed in terms of  the Kreft polynomial %$E^q b_{p/ q}(1/E)$ 
$ b_{p,q}(z)=-\sum_{j=0}^{\lfloor\frac{q}{2}\rfloor} a_{p,q}(2j)z^{2j}$
with  the Kreft coefficients \cite{Kreft} 
%\be \nonumber\det(m_{p/q}(e,0,0))+4(-1)^q=\sum_{j={q\over 2}-[{q\over 2}]}^{{q\over 2}}{\rm  c}_{p/q}%(2j)e^{2j}\ee
\begin{equation}\label{thea}\small{
a_{p,q}(2j)=(-1)^{j+1}\sum_{k_1=0}^{q-2j}\sum_{k_2=0}^{k_1}\ldots\sum_{k_{j}=0}^{k_{j-1}} 4\sin ^2\left(\frac{\pi  (k_1+2j-1) p}{q}\right)4\sin ^2\left(\frac{\pi  (k_2+2j-3) p}{q}\right)\ldots 4\sin ^2\left(\frac{\pi  (k_{j}+1) p}{q}\right)}
\end{equation}
 and
$a_{p,q}(0)=-1$. We refer   to e.g., \cite{Kreft}, and also  \cite{nous},  where details can be found on how to arrive at (\ref{thea}).
\iffalse
let us  denote the Kreft coefficient building blocks in  (\ref{building})  by ${{{\tilde{b}}}_{p/q}}(k)$ 
\be\nonumber
{{{\tilde{b}}}_{p/q}}(k) = 4\sin^2 \big( \frac{\pi k}{q} \big)
\equiv 2 - \rme^{\frac{2\rmi\pi k}{q}} - \rme^{-\frac{2\rmi\pi k}{q}} \;.
\label{fqk}
\ee
\fi
 In \cite{nous} we obtained { a closed expression for  the Hofstadter trace ${\rm Tr} H_{2\pi p/q}^{n}$  in terms of the Kreft coefficients (\ref{thea})}
 \be{\rm Tr} H_{2\pi p/q}^{n}= \frac{n}{q} \sum_{k \geq 0} \sum_{\substack{\ell_1,\ell_2,\ldots,\ell_{\lfloor q/2 \rfloor} \geq 0 \\ \ell_1 + 2\ell_2 + \cdots + \lfloor q/2 \rfloor \ell_{\lfloor q/2 \rfloor} = n/2 - kq}}
\frac{\binom{\ell_1+\ell_2 + \cdots + \ell_{\lfloor q/2 \rfloor} + 2k}{\ell_1,\ell_2,\ldots,\ell_{\lfloor q/2 \rfloor},2k}}{\ell_1+\ell_2 + \cdots + \ell_{\lfloor q/2 \rfloor} + 2k} \binom{2k}{k}^2\; \prod_{j = 1}^{\lfloor q/2 \rfloor} a_{p,q}(2j)^{\ell_j},\label{sumformula}
\ee
which in turn gave  $Z_n(e^{2i\pi p/q})$   via (\ref{thestuff}).  However both  (\ref{thea})  and  (\ref{sumformula}) are  somehow involved expressions which  cannot    be used in practice to reach the $C_n(A)$'s in (\ref{Zn_def}). 

The  observation  which allows for the lattice walks algebraic  area enumeration is that the $C_n(A)$'s are  contained in $[q] a_{p,q}(n)$ i.e.,
\begin{equation}\label{simplify} {1\over n}\sum_A C_{n}(A) e^{2iA\pi p/q} =[q] a_{p,q}(n)\end{equation}
where $[q] a_{p,q}(n)$  stands for the coefficient of the first order term in the $q$ expansion of the  Kreft coefficient $a_{p,q}(n)=q [q] a_{p,q}(n)+\ldots+q^{n/2}[q^{n/2}]a_{p,q}(n)$, a polynomial in $q$ of order $n/2$ with coefficients  which are { linear} combinations of  $\cos ({2 A \pi  p/q})$ with $A\in[0,\lfloor n^2/16\rfloor]$ ---see e.g., (\ref{a1q2weak}), (\ref{a4}), (\ref{a6})  and (\ref{a8}). 
This in turn implies that  the Hofstadter trace  simplifies to 
\begin{equation} \frac{1}{n} {\rm Tr} H_{2\pi p/q}^{n} = [q] a_{p,q}(n)\label{sonice}\end{equation} 
(how\footnote{Or  how to   reduce  (\ref{sumformula})  and  (\ref{partition}) ---see the Appendix--- to (\ref{sonice}).} to derive (\ref{simplify}) or equivalently (\ref{sonice})    will  be adressed elsewhere \cite{hard}.)

Likewise, the higher order terms in the $q$ expansion of  $a_{p,q}(n)$  are  given in terms of   $[q]a_{p,q}(n-2)$, $[q]a_{p,q}(n-4), \ldots$
\[a_{p,q}(2)=q[q]a_{p,q}(2)\]
 \[a_{p,q}(4)=q[q]a_{p,q}(4)-{q^2\over 2!}\left([q]a_{p,q}(2)\right)^2\] 
 \[a_{p,q}(6)=q[q]a_{p,q}(6)-q^2[q]a_{p,q}(2)[q]a_{p,q}(4)+{q^3\over 3!}\left([q]a_{p,q}(2)\right)^3\]
\iffalse \[ a_{p,q}(8)=q[q]a_{p,q}(8)-q^2[q]a_{p,q}(2)[q]a_{p,q}(6)-{q^2\over 2!}\left([q]a_{p,q}(4)\right)^2+{q^3\over 2!}\left([q]a_{p,q}(2)\right)^2[q]a_{p,q}(4)-{q^4\over 4!}\left([q]a_{p,q}(2)\right)^4\]\fi etc, i.e., \begin{equation}
 a_{p,q}(n)=-\sum_{k_j\ge 0\atop \sum_j jk_j=n/2} \prod_{j=1}^{n/2}(-1)^{k_j} \frac{1}{k_j !}\left(q[q]a_{p,q}(2j)\right)^{k_j}\nonumber\end{equation} which can be viewed, using (\ref{sonice}), 
\iffalse 
  \begin{equation}
 a_{p,q}(2j)=-\sum_{k_i\ge 0\atop \sum_i ik_i=j} \prod_{i=1}^{j}(-1)^{k_i} \frac{1}{k_i !}\left(\frac{q}{2i} {\rm Tr} H_{2\pi p/q}^{2i}\right)^{k_i}\nonumber\end{equation}\fi
  \begin{equation}
 a_{p,q}(n)=-\sum_{k_j\ge 0\atop \sum_j jk_j=n/2} \prod_{j=1}^{n/2}(-1)^{k_j} \frac{1}{k_j !}\left(\frac{q}{2j} {\rm Tr} H_{2\pi p/q}^{2j}\right)^{k_j}\nonumber\end{equation} as an inversion  of (\ref{sumformula}).
 
\iffalse  \begin{equation}
 a_{p,q}(2j)=-\sum_{k_i\ge 0\atop \sum_i ik_i=j} \prod_{i=1}^{j}(-1)^{k_i} \frac{1}{k_i !}\left([q]a_{p,q}(2i)\right)^{k_i}\label{sonicebis}\end{equation} \fi

 In the LHS of (\ref{simplify}) the algebraic area generating function $\sum_A C_{n}(A) e^{2iA\pi p/q}$ is defined  for all $n$ even and $q$. It follows that  in  its RHS  ---and in the equations below--- $a_{p,q}(n)$ should be understood as well  as  defined for all  $n$ even   and  $q$. However
the Kreft coefficient $a_{p,q}(2j)$   in (\ref{thea}) is not defined --in other words it trivially vanishes-- as soon as $q< 2j$. What is   meant in (\ref{simplify})  by   Kreft coefficient is the coefficient (\ref{thea}) defined for $q\ge 2j$ and  extrapolated  onto $q < 2j$ in such a way that it
obeys the same formula as for $q \ge 2j$, rather than trivially vanishing. This is all what is needed in view of the algebraic area enumeration. Still, for the sake of completeness, we  explain in the next section how to   explicitly build this extrapolation.

\section{Extrapolating  the Kreft coefficients}\label{Mash}
 % We set $p=1$ but all what follows  can be extended to $p$ coprime with $q$.
Let us denote by ${{{\tilde{b}}}_{p/q}}(k)$
 the building block $4\sin ^2\left(\frac{\pi  k p}{q}\right)=(1-e^{2 i k \pi  p \over q})(1-e^{-{2 i k\pi  p \over q}})$ appearing in  (\ref{thea})    so that  
     
\iffalse
\bea\nonumber
a_{p,q}(2j) 
 = (-1)^{j+1} \sum_{k_1=0}^{q-2j+1} \sum_{k_2=0}^{k_1} \cdots \sum_{k_{j}=0}^{k_{j-1}}
4 \sin^2\big( \frac{\pi(k_1+2j-1)}{q}\big)
4 \sin^2\big( \frac{\pi(k_2+2j-3)}{q}\big) \cdots
4 \sin^2\big( \frac{\pi(k_{j}+1)}{q}\big)\nonumber\\\label{apql}
\eea
\fi
\bea
a_{p,q}(2j) 
 = (-1)^{j+1} \sum_{k_1=0}^{q-2j+1} \sum_{k_2=0}^{k_1} \cdots \sum_{k_{j}=0}^{k_{j-1}}
{{\tilde{b}}}_{p/q}(k_1+2j-1)
{{\tilde{b}}}_{p/q}(k_2+2j-3) \cdots
{{\tilde{b}}}_{p/q}(k_{j}+1)\nonumber\\\label{apql}
\eea
where ${k_1=q-2j+1}$ has been added to the summation because it does not contribute anyway.
(\ref{apql}) is nonzero for $q \ge 2j-1$, otherwise the outermost sum trivially  vanishes by construction, in fact for  $q \ge 2j$, since  when $q=2j-1$, i.e., $k_1=0$, the outermost ${{\tilde{b}}}_{p/q}(k_1+2j-1)$ vanishes. 
 
%Let us first consider the simplest case $ a_{p,q}(2)$. 

\subsection{ Extrapolating $ a_{p,q}(2)= \sum_{k_1=0}^{q-1} {{\tilde{b}}}_{p/q}(k_1+1) $ }

\iffalse
we know  ({\em  Mathematica}  Simplify)
\be a_{p,q}(2)=2q\label{ach}\ee
However 
(\ref{apql})
\be\nonumber
a_{p,q}(2) = \sum_{k_1=0}^{q-1} {{\tilde{b}}}_{p/q}(k_1+1) 
\ee\fi
Shifting $k_1$ by $1$  one rewrites  $a_{p,q}(2)$  as
\bea
a_{p,q}(2)& = & \sum_{k_1=1}^{q} {{\tilde{b}}}_{p/q} (k_1) \nonumber\\
& = & 2q - \sum_{k_1=1}^{q} \big( \rme^{\frac{2\rmi k_1\pi p}{q}} +  \rme^{-\frac{2\rmi k_1\pi p}{q}} \big) \;. \label{a1q2}
\eea
For any $q > 1$, the sum in the second line of (\ref{a1q2}) vanishes, being the sum of the $q$-th roots of unity
of power $q$. However, for $q=1$ this sum reduces to a single term equal to 1+1.
Hence,
\bea\nonumber
a_{p,q}(2) & = & 2q \qquad {\rm when}\; q > 1 \;, \\\nonumber
a_{p,q}(2) & = & 0 \hspace{.2cm} \qquad {\rm when}\; q = 1 \;, 
\eea
which is nothing but saying,   accordingly to (\ref{thea}),  that $ a_{p,q}(2) $ trivially vanishes when $q=1$   ---whereas  it is equal to $2q$ when $q>1$.
The extrapolation amounts to extending the first equation for $q > 1$ onto the second one for $q=1$, i.e., for any $q\ge 1$ one should  end up  with
\be
a_{p,q}(2) = 2q\;.
\label{a1q2weak}
\ee
 For this to happen, it suffices to define
\be
\sum_{k=1}^{q} \rme^{\frac{2\rmi k \pi  p}{q}} = 0
\label{sum2}
\ee
for all $q$, including $q=1$. Substituting this into (\ref{a1q2}) yields (\ref{a1q2weak}).

To extend this scheme onto any $a_{p,q}(2j)$
  it is necessary
to generalize (\ref{sum2})  
%(as already said  it is actually 0 for any $q>1$, being a sum of $1$-th powers of $q$-th roots of unity;
%however, for $q=1$, it is  $1$)
 to
\be
\sum_{k=1}^{q} \rme^{\frac{2\rmi k \pi  p}{q}j} = 0
\label{sum}
\ee
for any $q\ge 1$ ---this sum   is actually 0 when $j$ is not a multiple of $q$, being a sum of $j$-th powers of $q$-th roots of unity;
however,  when $j$ is a multiple of $q$ it is equal to $q$.
One has then to express $a_{p,q}(2j)$ as combinations of sums of products of ${{{\tilde{b}}}_{p/q}}(k)$ and
use (\ref{sum}) to evaluate those sums.
Specifically,  as we have  seen,
\be
{1\over q}a_{p,q}(2) = {1\over q}\sum_{k=1}^{q} {{{\tilde{b}}}_{p/q}}(k) = 2 \;.
\label{sumfqk}
\ee
Also
\be
{1\over q}\sum_{k=1}^{q} {{{\tilde{b}}}_{p/q}}^2(k) = {1\over q}\bigg(\sum_{k=1}^{q}
\big( 2 - \rme^{\frac{2\rmi k \pi  p}{q}} - \rme^{-\frac{2\rmi k \pi  p}{q}} \big)
\big( 2 - \rme^{\frac{2\rmi k \pi p}{q}} - \rme^{-\frac{2\rmi k \pi  p}{q}} \big)\bigg)
= 6 \;,
\label{sumfqk2}
\ee
because when  the parentheses are opened and (\ref{sum}) is used, only those terms survive
where a constant, not an exponential, ends up being  summed. They are  $4+1+1=6$ such terms.

Quite generally, following this line of reasoning  one  gets 
\be\label{sumfqkn}
{1\over q}\sum_{k=1}^{q} {{{\tilde{b}}}_{p/q}}^j(k) =  {2j \choose j} \;,
\ee
 the equality  becoming a strict equality for $q > j$. Note that ${2j \choose j}$  is the number of closed lattice walks of length  $2j$  on a 1d lattice.

On the practical side one also remarks  
 that this is precisely what one gets  with {\em  Mathematica}   Simplify  acting on (\ref{apql}) when $2j=2$  with  output $ 2q$ or on ${1\over q} \sum_{k=1}^{q} {{{\tilde{b}}}_{p/q}}^j(k)$ with     output   ${2j \choose j}$. In the sequel we will use, when needed,  {\em Mathematica} Simplify to  check results   involving expressions with extrapolated  coefficients. 
 
 One stresses again that for $2j=2$ starting from (\ref{apql})  defined for $q\ge 2$, obtaining  $a_{p,q}(2) = 2q$  and  deciding   that this expression is valid  for all $q$ including $q=1$  is all what is needed in view of the algebraic area enumeration.

\subsection{ Extrapolating $ a_{p,q}(4)=-\sum_{k_1=0}^{q-3} \sum_{k_2=0}^{k_1}
{{{\tilde{b}}}_{p/q}}(k_1+3){{{\tilde{b}}}_{p/q}}(k_2+1) $ }
Let us illustrate this scheme  on $a_{p,q}(4)$. On the one hand acting with {\em  Mathematica} Simplify  on   $a_{p,q}(4)$ as defined in (\ref{apql})   yields the output 
\be a_{p,q}(4)=q \left(7+2 \cos (\frac{2 \pi  p}{q})-2
   q\right)\label{a4}\ee
 On the other hand one has, \iffalse from (\ref{apql})
\be
a_{p,q}(4) =  -\sum_{k_1=0}^{q-3} \sum_{k_2=0}^{k_1}
{{{\tilde{b}}}_{p/q}}(k_1+3){{{\tilde{b}}}_{p/q}}(k_2+1) \;;
\label{a1q4def}
\ee
or,\fi  
upon redefining the indices,
\be
a_{p,q}(4) = -\sum_{k_1=3}^{q} \sum_{k_2=1}^{k_1-2} {{{\tilde{b}}}_{p/q}}(k_1){{{\tilde{b}}}_{p/q}}(k_2) \;
\label{a1q4}
\ee
%so that  $q\ge 3$, if not (\ref{a1q4}) trivially vanishes.
Let us  adjust the sum limits so as to end up with, among others,  products
of pieces of the form $\sum_{k=1}^{q} {{{\tilde{b}}}_{p/q}}^j(k)$, which can then be directly evaluated
\iffalse, in the weak sense,\fi using (\ref{sumfqkn}). One rewrites  the double sum in (\ref{a1q4}) as
\be
\sum_{k_1=3}^{q} \sum_{k_2=1}^{k_1-2} {{{\tilde{b}}}_{p/q}}(k_1){{{\tilde{b}}}_{p/q}}(k_2) =\sum_{k_1=1}^{q} \sum_{k_2=1}^{k_1-2} {{{\tilde{b}}}_{p/q}}(k_1){{{\tilde{b}}}_{p/q}}(k_2) =
\sum_{k_1=1}^{q} {{{\tilde{b}}}_{p/q}}(k_1) \big( \sum_{k_2=1}^{k_1-1} {{{\tilde{b}}}_{p/q}}(k_2) - {{{\tilde{b}}}_{p/q}}(k_1-1) \big) \;.
\label{sumsum}
\ee
In the first term on the RHS of (\ref{sumsum}) 
one turns the triangular sum into a product of two sums, taking advantage
of the fact that the summand is symmetric with respect to $k_1$ and  $ k_2$  
\be\nonumber
\sum_{k_1=1}^{q} \sum_{k_2=1}^{k_1-1} {{{\tilde{b}}}_{p/q}}(k_1) {{{\tilde{b}}}_{p/q}}(k_2) = 
\sum_{k_1=1}^{q} \sum_{k_2=1}^{q} {{{\tilde{b}}}_{p/q}}(k_1) {{{\tilde{b}}}_{p/q}}(k_2) - \sum_{k_1=1}^{q} {{{\tilde{b}}}_{p/q}}^2(k_1) - \sum_{k_1=1}^{q} \sum_{k_2=k_1+1}^{q} {{{\tilde{b}}}_{p/q}}(k_1) {{{\tilde{b}}}_{p/q}}(k_2)\;.
\ee
The last term   is  the first one with the opposite sign
so
\be \nonumber\sum_{k_1=1}^{q} \sum_{k_2=1}^{k_1-1} {{{\tilde{b}}}_{p/q}}(k_1) {{{\tilde{b}}}_{p/q}}(k_2)={1\over 2}\bigg(\big(\sum_{k=1}^{q}  {{{\tilde{b}}}_{p/q}}(k)\big)^2 - \sum_{k=1}^{q} {{{\tilde{b}}}_{p/q}}^2(k)\bigg)\ee
and one arrives  at 
\begin{align}a_{p,q}(4)&= -\sum_{k_1=1}^{q} \sum_{k_2=1}^{k_1-2} {{{\tilde{b}}}_{p/q}}(k_1){{{\tilde{b}}}_{p/q}}(k_2) \label{nice}&=-{1\over 2}\bigg(\big(\sum_{k=1}^{q}  {{{\tilde{b}}}_{p/q}}(k)\big)^2 -\sum_{k=1}^{q} {{{\tilde{b}}}_{p/q}}^2(k) \bigg)+\sum_{k=1}^{q} {{{\tilde{b}}}_{p/q}}(k) {{{\tilde{b}}}_{p/q}}(k-1)\;.\end{align}
%{\bf End simpler  version.}
\iffalse
,   and  double checks that as it should
\begin{align}
\nonumber &=\sum_{l'=1}^q\sum_{k_2=1}^ {l'-1}  {{{\tilde{b}}}_{p/q}}(l'){{{\tilde{b}}}_{p/q}}(k_2)  -\sum_{l'=1}^{q}{{{\tilde{b}}}_{p/q}}(l') {{{\tilde{b}}}_{p/q}}(l'-1)\\\nonumber &=\sum_{l'=1}^q\sum_{k_2=1}^{ l'-2}  {{{\tilde{b}}}_{p/q}}(l'){{{\tilde{b}}}_{p/q}}(k_2) \end{align}

%\be \label{alas}a_{p,q}(4)=\sum_{l'={\it\bf 3}}^q\sum_{k_2=1}^{ l'-2}  {{{\tilde{b}}}_{p/q}}(l'){{{\tilde{b}}}_{p/q}}(k_2) \ee

\fi
For the first two terms in the RHS of  (\ref{nice}), use (\ref{sumfqk}) and (\ref{sumfqk2}), respectively;
hence,
\be
-{1\over 2}\bigg(\big(\sum_{k=1}^{q}  {{{\tilde{b}}}_{p/q}}(k)\big)^2 - \sum_{k=1}^{q} {{{\tilde{b}}}_{p/q}}^2(k) \bigg) = -{1\over 2}\big(4q^2 - 6q \big)\;.
\label{sumsumff}
\ee
For the last term, acting in the same way as in (\ref{sumfqk2}),
one gets
\bea
\sum_{k=1}^{q} {{{\tilde{b}}}_{p/q}}(k) {{{\tilde{b}}}_{p/q}}(k-1) & = & \sum_{k=1}^{q}
\big( 2 - \rme^{\frac{2\rmi\pi k p}{q}} - \rme^{-\frac{2\rmi\pi k p}{q}} \big)
\big( 2 - \rme^{\frac{2\rmi\pi (k-1)p}{q}} - \rme^{-\frac{2\rmi\pi (k-1) p}{q}} \big) \nonumber \\
& = & q\big(4 + \rme^{\frac{2\rmi\pi p}{q}} + \rme^{-\frac{2\rmi\pi p}{q}}\big)
= 2q\big(2 + \cos(\frac{2\pi p}{q})\big) \;,
\label{2term}
\eea
where, again, the parentheses have been opened and only those terms
where a constant ends up being summed have been kept.
\iffalse
; note the counting similarity between (\ref{sumfqk2}) and  (\ref{2term})
$$\sum_{k=1}^{q} {{{\tilde{b}}}_{p/q}}^2(k)  
= q\;\big( 6\big) $$
and 
$$\sum_{k=1}^{q} {{{\tilde{b}}}_{p/q}}(k) {{{\tilde{b}}}_{p/q}}(k-1) 
= q\;\big(4 + 2\cos(\frac{2\pi p}{q})\big).$$ 
\fi

Combining (\ref{nice}), (\ref{sumsumff})  and (\ref{2term})
one finally gets 
\be
a_{p,q}(4) = -2q^2 + 7q + 2q\,\cos(\frac{2\pi p}{q})
\label{a1q4res}
\ee
i.e.,  (\ref{a4}). An exact calculation starting from (\ref{apql}) or (\ref{nice})  would yield the same result
when  $q \ge 4$ but, trivially, zero when $q < 4$. But the RHS of (\ref{a1q4res}) does not trivially vanish 
when $q < 4$ ---it does vanish when $q=3$, but this is a non trivial vanishing ---see   (\ref{vanish}) in the Appendix. 

Once again, postulating (\ref{sum}) results in $a_{p,q}(4)$ getting
extrapolated onto $q < 4$.
In general for any $j$  the same approach should be used:
adjust the limits of sums, turn subdiagonal sums into products of independent sums,
and resolve the sums of products of ${{{\tilde{b}}}^{l_1}_{p/q}}(k)$, ${{{\tilde{b}}}_{p/q}}^{l_1}(k){{{\tilde{b}}}_{p/q}}^{l_2}(k-1)$, etc.

\section{ Algebraic area enumeration}

\subsection{At order $q$:  $[q]a_{p,q}(n)$ }
Focusing now on $a_{p,q}(4)$ at order $q$,  which is the part of interest  for the algebraic area  enumeration (\ref{simplify}),  one should discard  in (\ref{nice}) the   term  which is   a product of two sums since it contributes at order $q^2$. The two other terms    where only one sum appears   contribute at order $q$. They are   labelled by the $2^{4/2-1}=2$ compositions of $2$,  $2=2$ and $2=1+1$, 
\begin{align}\nonumber q[q]a_{p,q}(4)=&+{1\over 2} \sum_{k=1}^{q} {{{\tilde{b}}}_{p/q}}^2(k)\\ &+ \sum_{k=1}^{q} {{{\tilde{b}}}_{p/q}}(k) {{{\tilde{b}}}_{p/q}}(k-1)\label{4}\;.\end{align}
The algebraic area enumeration for walks of length $4$ has thus narrowed down to reduce  (\ref{a1q4}) to (\ref{4}) and to compute the two terms  in the  RHS of the latter.  This has been done as indicated above % and coincides with the {\em Mathematica}  Simplify  output
\begin{align}   \nonumber & {1\over q}\sum_{k=1}^{q} {{{\tilde{b}}}_{p/q}}^2(k)= 2(3)\\ \label{4bis} &{1\over q}\sum_{k=1}^{q} {{{\tilde{b}}}_{p/q}}(k) {{{\tilde{b}}}_{p/q}}(k-1)= 2\big(2+\cos(\frac{2 \pi  p}{q})\big)\end{align}
leading  to  the counting  $4/2\times 2(3)+4\times 2(2)=28$ walks with algebraic area $0$  and $4\times 2=8=4+4$ walks with algeabric area $\pm 1$.

{ This pattern is easily seen to  generalize to larger $n$'s as illustrated in the Appendix in the case $n=6$ and $n=8$}. 
Clearly  when  $n=4, 6, 8$ the algebraic area enumeration has narrowed down  to finding the coefficients   in front of  the $2^{n/2-1}$  terms in  (\ref{4}), (\ref{6}) and (\ref{8}), and  those of their $\cos({2k\pi p / q})$ expansions  in (\ref{4bis}), (\ref{6bis})  and (\ref{8bis}).

When $q=1$ the LHS of  (\ref{simplify})  counts   the number ${n \choose n/2}^2$ of closed walks on a square lattice of length $n$, up to a factor $1/n$.  One notes  in the three cases above $n=4, 6, 8$  that   
\begin{itemize}
\item
the coefficients in the cosine expansion  of each of the  $2^{n/2-1}$  terms  contributing  when $n=4$ to (\ref{4bis}),   when $n=6$ to (\ref{6bis})   and  when $n=8$ to (\ref{8bis}), add up   to ${4 \choose 2}$,  ${6 \choose 3}$  and  ${8 \choose 4}$ respectively. %  where ${4 \choose 2}^2$  does count the number of closed random walks of length 4 on a square lattice.
\item
when $q=1$  one  can  verify  the $n=4, 6, 8$ lattice walk countings  \be\label{4c}{4 \choose 2}(1/2+1)={4 \choose 2}^2/4\Rightarrow  1/2+1={4 \choose 2}/4\ee 
  \be\label{6c}{6 \choose 3}(1/3+1+1+1)={6 \choose 3}^2/6\Rightarrow 1/3+1+1+1={6 \choose 3}/6\ee 
  \be\label{8c}{8 \choose 4}(1/4+1+1+3/2+2+1+1+1)={8 \choose 4}^2/8\Rightarrow 1/4+1+1+3/2+2+1+1+1={8 \choose 4}/8\ee
  which hold because the coefficients in front of the said  terms  also add up to ${4 \choose 2}$ up to a factor $1/4$,  ${6 \choose 3}$ up to a factor $1/6$  and 
 ${8 \choose 4}$ up to a factor $1/8$ respectively.
\end{itemize}
\noindent The  countings (\ref{4c}, \ref{6c}, \ref{8c})
are particular cases of two general properties:  by construction 
\begin{itemize}
\item
the coefficients in the cosine expansion  of  each of the $2^{n/2-1}$  terms contributing to the first order $[q]a_{p,q}(n)$ add up  to ${n \choose n/2}$
\item
the coefficients  in front of the said   terms  also  add up  to  ${n\choose n/2}$ ---up to a $1/n$ factor.
\end{itemize}
${n\choose n/2}$ counts the number of closed  walks  of length $n$ on a 1d lattice. 
This hints to the fact  that both set of coefficients, those in front of the  terms contributing to  $[q]a_{p,q}(n)$ as well as those appearing in their cosine expansions,  might be expressable and interpretable in terms of properties  of 1d closed random walks of length $n$.

We now explain how to proceed\footnote{One remarks that   mirror symmetric compositions give identical enumerations, see for example  when $n=8$  the compositions $4=2+1+1$ and $4=1+1+2$   with the same coefficient 1 and same output
\begin{align*}\sum_{k=1}^q {{{\tilde{b}}}_{p/q}}^2(k)
   {{{\tilde{b}}}_{p/q}}(k-1) {{{\tilde{b}}}_{p/q}}(k-2)&=\sum_{k=1}^q {{{\tilde{b}}}_{p/q}}(k) {{{\tilde{b}}}_{p/q}}(k-1)
   {{{\tilde{b}}}_{p/q}}^2(k-2)\\&=2\big(12 + 14 \cos(2 \pi p/q) + 8 \cos(4 \pi p/q) + \cos(6 \pi p/q)\big)\;.\end{align*} 
   One could then   restrict to mirror-free compositions, { with all compositions   weighted  twice  except the  palindromic ones.  The number of mirror free compositions of $n/2$ is $(2^{n/2-1} +2^{\lfloor n/4\rfloor})/2$, the number of palindromic compositions being  $2^{\lfloor n/4\rfloor}$  }.}
 for a general $n$.  We stress that the  results to follow  will in part be based on experimental observations and deductions and not   actual derivations, which will remain to be built.

\subsection{The coefficients of the $2^{n/2-1}$  terms} 

\subsubsection{ Combinatorics}

For a given $n$, let us consider the  $2^{n/2-1}$ compositions of $n/2=l_1+l_2+\ldots+l_{n/2}$,   
$l_1\ge 1,\;l_2,\ldots, l_{n/2}\ge 0$.  { One infers from the cases and remarks above that } in the reduction at order $q$  of $a_{p,q}(n)$ 
these compositions label    the coefficients $c(l_1,l_2,\ldots,l_{n/2})$ of the $\sum _{k=1}^q {{{\tilde{b}}}_{p/q}}^{{l_1}}(k){{{\tilde{b}}}_{p/q}}^{l_2}(k-1)\ldots {{{\tilde{b}}}_{p/q}}^{l_{n/2}}(k-{n/2}+1)$'s, namely $[q]a_{p,q}(n)$  rewrites as
\be[q]a_{p,q}(n)=\sum_{l_1, l_2, \ldots, l_{n/2}\atop { \rm composition}\;{\rm of}\;n/2} c(l_1,l_2,\ldots,l_{n/2} ){1\over q}\sum _{k=1}^q {{{\tilde{b}}}_{p/q}}^{{l_1}}(k){{{\tilde{b}}}_{p/q}}^{l_2}(k-1)\ldots {{{\tilde{b}}}_{p/q}}^{l_{n/2}}(k-n/2+1)\label{labell}\ee
where 
\be\label{ouf} c(l_1,l_2,\ldots,l_{n/2})= \frac{{l_1+l_2\choose l_1}}{l_1+l_2}\;\; l_2\frac{{l_2+l_3\choose l_2}}{l_2+l_3}\;\ldots \;\; l_{{n/2}-1}\frac{{l_{{n/2}-1}+l_{n/2}\choose l_{{n/2}-1}}}{l_{{n/2}-1}+l_{n/2}}\ee
In (\ref{ouf})  $c(l_1,l_2,\ldots,l_{n/2})$ has been given in an explicitly  symmetric form $c(l_1,l_2,\ldots,l_{n/2})=c(l_{n/2},l_{{n/2}-1},\ldots,l_1)$, paying  attention to the fact that the last   block 
\be  l_{{n/2}-1}\frac{{l_{{n/2}-1}+l_{n/2}\choose l_{{n/2}-1}}}{l_{{n/2}-1}+l_{n/2}}\nonumber\ee
 which is equal to 1 when $l_{n/2}$ is equal to $0$, so that  $c(l_1,l_2,\ldots,l_{n/2})$  reduces to $c(l_1,l_2,\ldots,l_{n/2-1},0)$, should also be considered as equal to $1$  when both $l_{n/2}$ and $l_{n/2-1}$  are equal to $0$, these considerations extending to  $l_{n/2}$, $l_{n/2-1}$ and $l_{n/2-2}$ equal to 0, etc. 
 
\noindent One also checks that 
\be \sum_{l_1, l_2, \ldots, l_{n/2}\atop {\rm composition}\;{\rm of}\;n/2}c(l_1,l_2,\ldots,l_{n/2})=\frac{{n\choose {n/2}}}{n}\; \label{right}\ee
  as it should.

\iffalse 
\begin{align*}2j=10 :& \;{10 \choose 5}(1/5+1+1+2+2+3+1+1+3+3+1+2+2+1+1+1)={10 \choose 5}^2/10\\\Rightarrow \;&{10 \choose 5}/2=1+10+20+15+10+30+5+20+10+5\end{align*}
\fi

%\subsubsection 

\subsubsection{ Interpretation in terms of 1d lattice walk counting}
 It is sufficient to restrict to  closed 1d lattice  walks of length $n$  making the first step to the  right: they are   ${n\choose {n/2}}/2$  such walks each made of $n/2$ right steps and $n/2$ left steps, or, equivalently, $n/2$ right-left steps. Let us focus on the number of right-left steps  which appear on top of each other  along the  $n$ steps made by the lattice walk.

\noindent For e.g., $n=6$   the coefficients in \be\nonumber\sum _{k=1}^q {{{\tilde{b}}}_{p/q}}^3(k)+3\sum _{k=1}^q 
   {{{\tilde{b}}}_{p/q}}^2(k) {{{\tilde{b}}}_{p/q}}(k-1)+3\sum _{k=1}^q 
   {{{\tilde{b}}}_{p/q}}(k) {{{\tilde{b}}}_{p/q}}^2(k-1)+3\sum _{k=1}^q {{{\tilde{b}}}_{p/q}}(k) {{{\tilde{b}}}_{p/q}}(k-1){{{\tilde{b}}}_{p/q}}(k-2)\;\ee
   i.e., $1+3+3+3={6\choose 3}/2$, count  1 lattice walk  with the 3 right-left steps on top of each other ${{{\tilde{b}}}_{p/q}}^3(k)$, 3 lattice walks with 2 right-left steps on top of each other followed by 1 right-left step $ 
   {{{\tilde{b}}}_{p/q}}^2(k) {{{\tilde{b}}}_{p/q}}(k-1)$,  3 lattice walks with 1 right-left step followed by 2 right-left steps on top of each other $ 
   {{{\tilde{b}}}_{p/q}}(k) {{{\tilde{b}}}_{p/q}}^2(k-1)$ and 3 lattice walks  with  3 right-left steps following each other ${{{\tilde{b}}}_{p/q}}(k) {{{\tilde{b}}}_{p/q}}(k-1){{{\tilde{b}}}_{p/q}}(k-2)$ ---see Figure 1.
   
   \begin{figure}
%\begin{center}
%\hspace{2cm}
%\vspace{2cm}
\includegraphics[scale=.7]{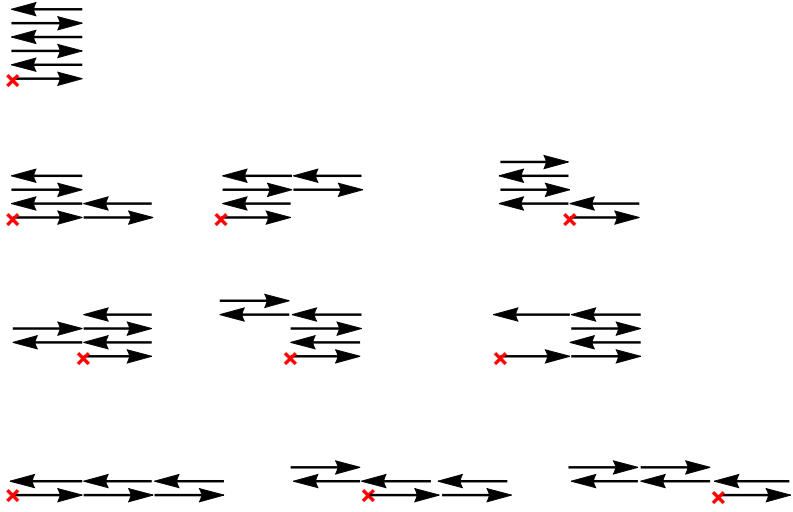}\label{nis6}
\caption{$n=6$ : the 10 lattice walks counted as "1+3+3+3=10". The walks have been spread in the vertical direction to facilitate the visualization of  the counting. The cross denotes the starting and ending point of each walk.}
%\end{center}
\end{figure}

  \noindent  For e.g.,  $n=8$  the coefficients in 
  \begin{align*}&\sum _{k=1}^q {{{\tilde{b}}}_{p/q}}^4(k)+4\sum _{k=1}^q 
   {{{\tilde{b}}}_{p/q}}^3(k) {{{\tilde{b}}}_{p/q}}(k-1)+4\sum _{k=1}^q 
   {{{\tilde{b}}}_{p/q}}(k) {{{\tilde{b}}}_{p/q}}^3(k-1)+6\sum _{k=1}^q {{{\tilde{b}}}_{p/q}}^2(k) {{{\tilde{b}}}_{p/q}}^2(k-1)+\\&8\sum _{k=1}^q {{{\tilde{b}}}_{p/q}}(k) {{{\tilde{b}}}_{p/q}}^2(k-1){{{\tilde{b}}}_{p/q}}(k-2)+4\sum _{k=1}^q {{{\tilde{b}}}_{p/q}}^2(k) {{{\tilde{b}}}_{p/q}}(k-1){{{\tilde{b}}}_{p/q}}(k-2)+\\&4\sum _{k=1}^q {{{\tilde{b}}}_{p/q}}(k) {{{\tilde{b}}}_{p/q}}(k-1){{{\tilde{b}}}_{p/q}}^2(k-2)+4\sum _{k=1}^q {{{\tilde{b}}}_{p/q}}(k) {{{\tilde{b}}}_{p/q}}(k-1){{{\tilde{b}}}_{p/q}}(k-2){{{\tilde{b}}}_{p/q}}(k-3)\;\end{align*} 
   i.e.,   $ 1+4+4+6+8+4+4+4={8\choose 4}/2$,
   count 1 lattice walk  with the 4 right-left  steps on top of each other ${{{\tilde{b}}}_{p/q}}^4(k)$,  4 lattice walks with 3 right-left steps on top of each other followed by 1 right-left step ${{{\tilde{b}}}_{p/q}}^3(k) {{{\tilde{b}}}_{p/q}}(k-1)$, 4 lattice walks with 1 right-left step followed by 3 right-left steps on top of each other ${{{\tilde{b}}}_{p/q}}(k) {{{\tilde{b}}}_{p/q}}^3(k-1)$, 6 lattice walks with 2 right-left steps on top of each other followed by 2 right-left steps on top of each other ${{{\tilde{b}}}_{p/q}}^2(k) {{{\tilde{b}}}_{p/q}}^2(k-1)$, etc.
   
   Quite generally the coefficient $c(l_1,l_2,\ldots,l_{n/2})$ in (\ref{labell})  with $l_1, l_2, \ldots, l_n/2$ a composition of $n/2$ does  count, when multiplied by $n$, the number of closed   lattice walks of length $n$ with $ l_1$ right-left  steps on top of each other  followed by $l_2$ right-left  steps on top of each other $\ldots$  followed by $l_{n/2}$ right-left  steps on top of each other.

\subsection{The coefficients in the cosine  expansions   %(a $1/q$  factor is always understood in front of each sum $\sum _{k=1}^q \ldots $)
} 

Combinatorial expressions for the coefficients of the cosine expansions of  $\sum _{k=1}^q {{{\tilde{b}}}_{p/q}}^{{l_1}}(k){{{\tilde{b}}}_{p/q}}^{l_2}(k-1)\ldots {{{\tilde{b}}}_{p/q}}^{l_{n/2}}(k-n/2+1)$ in (\ref{labell}) can be { seen in simple cases} to rewrite in term of products of deformed 1d  lattice  binomials ${2l_i\choose l_i}$  such as
   \begin{align} {1\over q}\sum _{k=1}^q {{{\tilde{b}}}_{p/q}}^{{l_1}}(k)={2l_1\choose l_1 }\label{lbisbis}\end{align}
 \begin{align}& {1\over q}\sum _{k=1}^q {{{\tilde{b}}}_{p/q}}^{{l_1}}(k){{{\tilde{b}}}_{p/q}}^{{l_2}}(k-1)=\nonumber\\&\sum_{A=-\infty}^{\infty}\cos\left({2A\pi p\over q}\right){2{l_1}\choose {l_1}+A}{2 {l_2} \choose {l_2}-A }\;,\label{llbisbis}\end{align}
 \begin{align}&{1\over q}\sum _{k=1}^q {{{\tilde{b}}}_{p/q}}^{{l_1}}(k){{{\tilde{b}}}_{p/q}}^{{l_2}}(k-1){{{\tilde{b}}}_{p/q}}^{{l_3}}(k-2)=\nonumber\\&\sum_{A= -\infty}^{ 
   \infty}\cos\left({2A\pi p\over q}\right)\sum_{k_3=-\infty}^{\infty}{2 l_1\choose l_1 - k_3 } {2 l_2\choose
      l_2 + 2 k_3 + A} {2 l_3\choose l_3 - k_3-A}\;, \label{lllbisbis}\end{align}
\begin{align}&{1\over q}\sum _{k=1}^q {{{\tilde{b}}}_{p/q}}^{{l_1}}(k){{{\tilde{b}}}_{p/q}}^{{l_2}}(k-1){{{\tilde{b}}}_{p/q}}^{{l_3}}(k-2){{{\tilde{b}}}_{p/q}}^{{l_4}}(k-3)=\nonumber\\&\sum_{A=-\infty}^{\infty}\cos\left({2A\pi p\over q}\right)\sum_{k_3= -\infty}^{ 
   \infty}\sum_{k_4= -\infty}^{ \infty}  {2 l_1\choose l_1 - k_4  - k_3} {2 l_2\choose 
     l_2 +k_4+2k_3} {2 l_3\choose l_3 +  k_4-k_3 + A} {2 l_4\choose 
     l_4 - k_4 -   A}  \;, \label{llllbisbis}\end{align}
\iffalse  \scriptsize 
    \begin{align}&{1\over q}\sum _{k=1}^q {{{\tilde{b}}}_{p/q}}^{{l_1}}(k){{{\tilde{b}}}_{p/q}}^{{l_2}}(k-1){{{\tilde{b}}}_{p/q}}^{{l_3}}(k-2){{{\tilde{b}}}_{p/q}}^{{l_4}}(k-3){{{\tilde{b}}}_{p/q}}^{{l_5}}(k-4)=\nonumber\\&\sum_{A=-\infty}^{\infty}\cos\left({2A\pi p\over q}\right)\sum_{k_3= -\infty}^{ 
   \infty}\sum_{k_4= -\infty}^{ \infty} \sum_{k_5= -\infty}^{ \infty} {2 l_1\choose l_1 - k_5 -   k_4-k_3 } {2 l_2\choose l_2 -  k_5+k_4+ 2k_3} {2 l_3\choose 
     l_3  + 4k_5+k_4-k_3} {2 l_4\choose l_4 - k_5 -k_4 + A } {2 l_5\choose 
     l_5- k_5-A  }  \label{lllllbisbis}\end{align}
     \normalsize \fi
     which can be checked to be, as it should,   $l_1,l_2,\ldots,l_j\to l_j,l_{j-1},\ldots,l_1$    symmetric ---e.g., in (\ref{llllbisbis}) one redefines $k_4=k'_4-k_3-A$ followed by $A\to -A$ and $k_3\to -k_3$. 
        
One { observes}  that in (\ref{llbisbis}-\ref{llllbisbis}), \iffalse \ref{lllllbisbis}), as well as in (\ref{1000}, \ref{1001}, \ref{1002}, \ref{1003}),\fi 
$+A$  and $-A$ always enter  in the last  two binomials respectively. 
With each additionnal  binomial an additionnal summation enters
which preserves  the  summations  already present. The binomials  have    the following $k_3, k_4, \ldots, k_j$ weight pattern  
\be\nonumber\begin{matrix}
 &\; &\;l_1 & \; & \; & \; & \; & \; & \; &\; \\
 %&\;  &\;   & \; & \; & \; & \; & \; & \; &\;\\
&\;\;\;\;\;&\;l_1 & \;l_2 &\;& \; & \; & \; & \; & \;\\
%&&\; &  \; &\;& \; & \; & \; & \; & \;\\
& & \;l_1&\;l_2&\;l_3 &\;& \; & \; & \;& \;  \\
 k_3&\;\;\;\;\;& -1\;&+2\;&-1  &\;& \; & \; & \;& \; \;\\
& \;\; &\;l_1 & \;l_2 &\;l_3&\;l_4&\;& \;  &\; & \; \\
k_4 &\;\;\;\;\;& -1 \;& +1 \;& +1\;&-1\; &\;& \;  &\; & \; \\
&& \;l_1& \;l_2&\;l_3 &\;l_4 &\;l_5&\;& \; & \;  \\
k_5& \;\;\;\;\;& -1\; & -1\; &+4\;&-1\; &-1&\;& \; & \;  \;\\
&& \;l_1 &\;l_2 &\;l_3&\;l_4&\;l_5&\;l_6  &\;&\; \\
k_6&\;\; \;\;\;& -1\; &-1\; &+2 \;& +2\;& -1\;& -1 \;\;\; &\;&\; \;\\
&&\;l_1 &\;l_2 &\;l_3 &\;l_4 &\;l_5&\;l_6&\;l_7&\;\\
k_7& \;\;\;\;\;& -1\; & -1\;  & -1\; & +6\; & -1\;&-1\;&-1\;\;\; &\; \;\\
& \; \;&\;l_1& \;l_2 & \;l_3  & \;l_4 &\;l_5&\;l_6 &\;l_7&\;l_8 \\
k_8& \;\;\;\;\;&-1\;&-1\; &-1\;  &+3\; &+3 \;&-1\; &-1\;&-1\; \\
\end{matrix}\;\ee
etc. One sees that for a given $k_j$  with $j\ge 3$
\begin{itemize}
\item
 the weights are symmetric and
 add up to $0$
\item
 when $j$ is even :  $j-2$  weights equal to $-1$ and $2$ weights  at the center  equal to $j/2-1$
\item
 when $j$ is odd :  $j-1$  weights equal to $-1$ and $1$ weight  at the center  equal to $j-1$.
\end{itemize}
One also notes that the $A$ summation  upper  and lower  limits are opposite and      bounded  by $\lfloor (l_1+l_2+\ldots+l_j)^2/4\rfloor$  and  $-\lfloor (l_1+l_2+\ldots+l_j)^2/4\rfloor$ respectively:  for example in (\ref{llbisbis}) and (\ref{1000})  the  actual upper limit of summation is $\min{(l_1,l_2)}\le \lfloor(l_1+l_2)^2/4\rfloor$; likewise  in (\ref{lllbisbis}) and (\ref{1001}) the actual upper limit of summation  can be found to be $l_1+l_2+l_3-\max{(|l_1-l_3|,l_2)}\le \lfloor(l_1+l_2+l_3)^2/4\rfloor$. It follows that in (\ref{llbisbis}-\ref{llllbisbis}) the summation  $\sum_{A=-\infty}^{\infty}$ means  at most $\sum_{A=-\lfloor (l_1+l_2+\ldots+l_j)^2/4\rfloor}^{\lfloor (l_1+l_2+\ldots+l_j)^2/4\rfloor}$ ---it is actually so  when  $l_1= l_2= \ldots= l_j= 1$, i.e., when the $l_i$'s do label a composition of $j$,  the  one whose  parts are all equal to $1$.   

One infers in general  that for $j\ge 2$   ---for $j=1$ { see (\ref{lbisbis})}

\begin{align} &{1\over q}\sum _{k=1}^q {{{\tilde{b}}}_{p/q}}^{{l_1}}(k){{{\tilde{b}}}_{p/q}}^{l_2}(k-1)\ldots {{{\tilde{b}}}_{p/q}}^{l_{j}}(k-j+1)=\nonumber\\&\sum_{A=-\infty}^{\infty}\cos\left({2A\pi p\over q}\right)\sum_{k_3= -\infty}^{ 
   \infty}\sum_{k_4= -\infty}^{ \infty}\ldots \sum_{k_j= -\infty}^{ \infty} \prod_{i=1}^j{2l_i\choose l_i-k_{i,j} +A(\delta_{i,j-1}-\delta_{i,j}%+\delta_{1,j}
   )}\label{matrix}\end{align}
   with    for $1\le i\le j$\iffalse 
   \footnote{It can be rewritten as \begin{align*}
k_{i,j}=-\sum_{r=2i+1}^j k_r-\sum_{r=3+s}^{j-s}\,\sum_{s=0}^{j-2}\delta _{i,r}k_{r+s}\,+\sum_{r=3+s}^{j+s-2[\frac{s}{2}]}\,\sum_{s=0}^{j-2}\delta _{i,2+[\frac{r-3}{2}]}k_{r-s+2[s/2]}
\;.\end{align*}}

\be k_{i,j}=-\sum_{r=i}^{2i-3}k_r+(i-2)k_{2i-2}+2(i-1)k_{2i-1}+(i-1)k_{2i}-\sum_{r=2i+1}^{j}k_r \label{welll}\ee \fi 
\be k_{i,j}=-\sum_{r=i}^{j}k_r+(i-1)k_{2i-2}+(2i-1)k_{2i-1}+ i k_{2i} \label{welll}\ee
   where  the  $k_{2i-2}, k_{2i-1}$ or $k_{2i}$ terms  do materialize  if  $2i-2$, $2i-1$ or $2i$ are  lower or equal to $j$ respectively. One notes that the $k_1$ and $k_2$ terms in (\ref{welll})  cancel out. Also the fact that when $l_j=0$ the RHS of (\ref{matrix})  does reduce  to %\[\sum_{A=-\infty}^{\infty}\cos\left({2A\pi p\over q}\right)\sum_{k_3= -\infty}^{ 
 %  \infty}\sum_{k_4= -\infty}^{ \infty}\ldots \sum_{k_{j-1}= -\infty}^{ \infty} \prod_{i=1}^{j-1}{2l_i\choose l_i-k_{i,j-1} +A(\delta_{i,j-2}-\delta_{i,j-1})}\] 
 the same form with $j$  replaced by $j-1$ can be seen  via  some appropriate $k_i$ redefinitions.

When $q=1$,  
     (\ref{llbisbis}-\ref{llllbisbis}) \iffalse, \ref{lllllbisbis})\fi  reduce  { as it should  to the binomial countings}
    \begin{align} &{2(l_1+l_2)\choose l_1+l_2}=\sum_{A=-\infty}^{\infty}{2{l_1}\choose {l_1}+A}{2 {l_2} \choose {l_2}-A }\label{1000}\end{align}
    \begin{align}&{2(l_1+l_2+l_3)\choose l_1+l_2+l_3}=\sum_{A=-\infty}^{\infty}\sum_{k_3=-\infty}^{\infty}{2 l_1\choose l_1 - k_3} {2 l_2\choose
      l_2 + 2 k_3 + A} {2 l_3\choose l_3 - k_3 - A}\label{1001}\end{align}
 \begin{align}&{2(l_1+l_2+l_3+l_4)\choose l_1+l_2+l_3+l_4}\nonumber\\=&
   \sum_{A=-\infty}^{\infty}\sum_{k_3= -\infty}^{ 
   \infty}\sum_{k_4= -\infty}^{ \infty}  {2 l_1\choose l_1 - k_4 -   k_3} {2 l_2\choose 
     l_2 + k_4 + 2 k_3} {2 l_3\choose l_3 +k_4-k_3 + A} {2 l_4\choose 
     l_4 - k_4  - A}\label{1002}\end{align}
  \iffalse
  \scriptsize \begin{align}&{2(l_1+l_2+l_3+l_4+l_5)\choose l_1+l_2+l_3+l_4+l_5}=\nonumber\\&\sum_{A=-\infty}^{\infty}\sum_{k_3= -\infty}^{ 
   \infty}\sum_{k_4= -\infty}^{ \infty} \sum_{k_5= -\infty}^{ \infty} {2 l_1\choose l_1 - k_5-   k_4-k_3 } {2 l_2\choose l_2 -k_5+k_4+2k_3} {2 l_3\choose 
     l_3 + 4k_5+k_4-k_3} {2 l_4\choose l_4 - k_5 -k_4 + A} {2 l_5\choose 
     l_5 - k_5-A }  \label{1003}  \end{align}\normalsize\fi
    and in general (\ref{matrix}) reduces   to the counting
   \begin{align}\label{muiz} &{2(l_1+l_2+\ldots+l_j)\choose l_1+l_2+\ldots+l_j}=\nonumber\\&\sum_{A=-\infty}^{\infty}\sum_{k_3= -\infty}^{ 
   \infty}\sum_{k_4= -\infty}^{ \infty}\ldots \sum_{k_j= -\infty}^{ \infty} \prod_{i=1}^j{2l_i\choose l_i-k_{i,j} +A(\delta_{i,j-1}-\delta_{i,j}%+\delta_{1,j}
   )}\end{align}
   where, as well as in (\ref{matrix}),  the summation $\sum_{A=-\infty}^{\infty}$ is understood as at most $\sum_{A=-\lfloor (l_1+l_2+\ldots+l_j)^2/4\rfloor}^{\lfloor (l_1+l_2+\ldots+l_j)^2/4\rfloor}$.
   
    {  The binomial counting  (\ref{muiz}) can be easily checked by first summing over $A$  using (\ref{use}),  redefining the $k_i$'s such that one arrives at   the same expression now  for  $l_{j-1}+l_j,  l_{j-2}, \ldots, l_1$, i.e., by symmetry for $l_1, \ldots,  l_{j-2},l_{j-1}+ l_j$, and then  repeating the procedure $j-2$ times to  obtain the  binomial ${2(l_1+l_2+\ldots+l_j)\choose l_1+l_2+\ldots+l_j}$. For example in the case  $j=5$ after summing over  $A$  one redefines $2k_5+k_4=k'_4+k'_3$, $4k_5+k_4-k_3=k'_4+2k'_3$  and $k_5=A$, and so on.}
   
   One notes that an interpretation, if any, of the coefficients  in (\ref{matrix})  in terms of  1d lattice walks   is lacking.

\section{Conclusion}
The algebraic area enumeration for lattice random walks follows from (\ref{labell}), (\ref{ouf})  and (\ref{matrix}). Using (\ref{simplify}) one { can then conjecture}    that  for $n\ge 4$  (for $n=2$ trivially $C_2(A)=4 \delta_{A,0}$)
\begin{align} C_{n}(A) =&\; n \sum_{l_1, l_2, \ldots, l_{n/2}\atop{\rm composition}\;{\rm of}\; n/2} \frac{{l_1+l_2\choose l_1}}{l_1+l_2}\;\; l_2\frac{{l_2+l_3\choose l_2}}{l_2+l_3}\;\ldots \;\; l_{{n/2}-1}\frac{{l_{{n/2}-1}+l_{n/2}\choose l_{{n/2}-1}}}{l_{{n/2}-1}+l_{n/2}}\nonumber\\& \sum_{k_3= -\infty}^{ 
   \infty}\sum_{k_4= -\infty}^{ \infty}\ldots \sum_{k_{n/2}= -\infty}^{\infty}\prod_{i=1}^{n/2}{2l_i\choose l_i-k_{i,n/2} +A(\delta_{i,n/2-1}-\delta_{i,n/2}%+\delta_{1,n/2}
   )}\;.\label{theformula}\end{align}\iffalse
   { One can  check that $C_{n}(A)$  vanishes when $A$ is not in between  $-\lfloor n^2/16\rfloor$ and $\lfloor n^2/16\rfloor$, that $C_n(A)$ reduces when  $A=\pm\lfloor n^2/16\rfloor$  to $n$  when $n/2$ is even and $2n$ when $n/2$  is odd   and that the lattice walks counting  $\sum_{A}C_{n}(A)={n\choose n/2}^2$ holds using (\ref{muiz}) and then (\ref{right}).  }\fi{ One  checks that $C_{n}(A)$  vanishes when $A$ is not in between  $-\lfloor n^2/16\rfloor$ and $\lfloor n^2/16\rfloor$ and that the lattice walks counting  $\sum_{A}C_{n}(A)={n\choose n/2}^2$ holds by using (\ref{muiz})  and then (\ref{right}).  }

The complexity of the formula (\ref{theformula}) increases  quickly with $n$ { since the  number of compositions grows exponentially with $n$}.   We have verified (\ref{theformula}) for small $n$  against complete enumeration 
\iffalse up to $n=12$ \fi ---see e.g., \cite{mash} for a complete enumeration based on a recurrence relation  which encodes the algebraic area combinatorics; see also \cite{epelbaum}  for recent efforts on the algebraic area enumeration of lattice walks.  Larger $n$  verifications   would gain in  a better understanding  of bounds on  the $k_3, k_4, \ldots$  opposite upper and lower  limits ---clearly in (\ref{lllbisbis}) and (\ref{1001}) the $k_3$ summation goes from  $-\min{(l_1,l_2+l_3)}$ to $ \min{(l_1,l_2+l_3)}$; in (\ref{llllbisbis}) and (\ref{1002})  from $-\min{(l_1+l_2,l_3+l_4)}$ to $ \min{(l_1+l_2,l_3+l_4)}$, etc\footnote{ One can rewrite the multiple sum in (\ref{theformula}) in the less symmetric form   \begin{align}C_{n}(A) =&\; n \sum_{l_1, l_2, \ldots, l_{n/2}\atop{\rm composition}\;{\rm of}\; n/2} \frac{{l_1+l_2\choose l_1}}{l_1+l_2}\;\; l_2\frac{{l_2+l_3\choose l_2}}{l_2+l_3}\;\ldots \;\; l_{{n/2}-1}\frac{{l_{{n/2}-1}+l_{n/2}\choose l_{{n/2}-1}}}{l_{{n/2}-1}+l_{n/2}}\nonumber \\&\sum_{k_3= 0}^{2l_3 
 }\sum_{k_4= 0}^{ 2l_4}\ldots \sum_{k_{n/2}= 0}^{2l_{n/2}}{2l_1\choose l_1 +A+\sum_{i=3}^{n/2}(i-2)(k_i-l_i)}{2l_2\choose l_2 -A-\sum_{i=3}^{n/2}(i-1)(k_i-l_i)}\prod_{i=3}^{n/2}{2l_i\choose k_i}\label{trans}\end{align} but with explicit upper and lower summation limits. }. 
 
 { An opened issue concerns the $n\to\infty$ limit  in (\ref{theformula}), where one should,  with an appropriate vanishing  lattice spacing scaling, recover L\'evy's  Brownian  law (\ref{Levy}).}

We  note  that altogether with (\ref{Zn_def}) and (\ref{thestuff}), the algebraic area enumeration (\ref{theformula})  gives in the commensurate case a combinatorial expression for the $n$-th moment  of the Hofstafdter Hamiltonian\be\sum_A C_{n}(A) e^{2iA\pi p/q}={\rm Tr} H_{2\pi p/q}^{n}\nonumber\;.\ee\iffalse
\begin{align*} {\rm Tr} H_{2\pi p/q}^{n}=&\;n\sum_A e^{2iA\pi p/q} \sum_{{\rm compositions}\;{\rm of}\; n/2} \frac{{l_1+l_2\choose l_1}}{l_1+l_2}\;\; l_2\frac{{l_2+l_3\choose l_2}}{l_2+l_3}\;\ldots \;\; l_{{n/2}-1}\frac{{l_{{n/2}-1}+l_{n/2}\choose l_{{n/2}-1}}}{l_{{n/2}-1}+l_{n/2}}\nonumber\\& \sum_{k_3= -\infty}^{ 
   \infty}\sum_{k_4= -\infty}^{ \infty}\ldots \sum_{k_{n/2}= -\infty}^{\infty}\prod_{i=1}^j{2l_i\choose l_i-s_{i,n/2} +A(\delta_{i,n/2-1}-\delta_{i,n/2})}\end{align*}\fi 
The irrational limit, where both $p$ and $q$ go to infinity,   amounts   to directly  trade    $2\pi p/q$ for $\gamma$ in $ e^{2iA\pi p/q}$.
It would be  interesting to see if any  insights   on this limit are gained by doing so. 

{ Finally  in the Appendix we  generalize  $C_{n}(A) $ in (\ref{theformula})  to $C_{m,m,n/2-m,n/2-m}(A) $, the number  of closed lattice  walks of length $n$ with $m$ steps  right, $m$ steps left, $n/2-m$ steps up, $n/2-m$ steps  down  enclosing the algebraic area $A$. }

  \section*{Acknowledgements}

S.O. would like to thank Alain Comtet, Olivier Giraud, Alexios Polychronakos  and Stephan Wagner for interesting conversations.  Special thanks to Stefan Mashkevich for his precious help  in  Section \ref{Mash}.

%\pagebreak

\section*{Appendix }

\subsection*{At order q: $[q]a_{p,q}(6)$ and $[q]a_{p,q}(8)$}

The same logic at work for $a_{p,q}(4)$ should prevail in the case  $n=6$ with  {\em  Mathematica}   output
\be\label{a6}a_{p,q}(6)=\frac{2}{3}  q \left(58 + 36\cos (\frac{2 \pi  p}{q})+ 6 \cos (\frac{4 \pi  p}{q})- q \big(21 + 6 \cos (\frac{2 \pi  p}{q})\big) +2 q^2\right)\;.\ee
 One has from  (\ref{apql})
\begin{align}
      \nonumber  a_{p,q}(6)=\sum_{k_1=0}^{q-5}\sum_{k_2=0}^{k_1}\sum_{k_3=0}^{k_2}  {{{\tilde{b}}}_{p/q}}(k_1+5){{{\tilde{b}}}_{p/q}}(k_2+3){{{\tilde{b}}}_{p/q}}(k_3+1) \end{align}
i.e., upon redifining the indices,
\begin{align}\label{nicenicebis}
a_{p,q}(6)=\sum_{k_1=5}^q\sum_{k_2=3}^{k_1-2}\sum_{k_3=1}^{k_2-2}  {{{\tilde{b}}}_{p/q}}(k_1){{{\tilde{b}}}_{p/q}}(k_2){{{\tilde{b}}}_{p/q}}(k_3)=\sum_{k_1=1}^q\sum_{k_2=1}^{k_1-2}\sum_{k_3=1}^{k_2-2}  {{{\tilde{b}}}_{p/q}}(k_1){{{\tilde{b}}}_{p/q}}(k_2){{{\tilde{b}}}_{p/q}}(k_3)\;. \end{align}
%so that $q\ge 5$, if not it  trivially vanishes.
Similarly 
   in the reduction of   (\ref{nicenicebis}) at order q  only the terms   with only one sum  contribute:  there are four such terms labelled by the $2^{6/2-1}=4$ compositions of $3$,  $3=3$, $3=2+1$, $3=1+2$ and $3=1+1+1$
\begin{align} q[q]a_{p,q}(6) =&+\frac{1}{3} \sum _{k=1}^q {{{\tilde{b}}}_{p/q}}^3(k)\nonumber\\&+ \sum _{k=1}^q 
   {{{\tilde{b}}}_{p/q}}^2(k) {{{\tilde{b}}}_{p/q}}(k-1)\nonumber\\&+
 \sum _{k=1}^q
  {{{\tilde{b}}}_{p/q}}(k) {{{\tilde{b}}}_{p/q}}^2(k-1)\nonumber \\&+ \sum _{k=1}^q {{{\tilde{b}}}_{p/q}}(k) {{{\tilde{b}}}_{p/q}}(k-1){{{\tilde{b}}}_{p/q}}(k-2)\label{6}\end{align}
  with    output
  \begin{align}\nonumber &{1\over q}\sum _{k=1}^q {{{\tilde{b}}}_{p/q}}^3(k)=4\big(5\big)\\&\nonumber{1\over q}\sum _{k=1}^q 
   {{{\tilde{b}}}_{p/q}}^2 (k){{{\tilde{b}}}_{p/q}}(k-1)=4\big(3+ 2\cos({2 \pi p\over q})\big)\\&\nonumber
{1\over q}\sum _{k=1}^q
  {{{\tilde{b}}}_{p/q}}(k) {{{\tilde{b}}}_{p/q}}^2(k-1) =4\big(3+ 2\cos({2 \pi p\over q})\big)\\&\label{6bis}{1\over q}\sum _{k=1}^q {{{\tilde{b}}}_{p/q}}(k) {{{\tilde{b}}}_{p/q}}(k-1){{{\tilde{b}}}_{p/q}}(k-2)=4\big(2+2\cos({2 \pi p\over q})+\cos({4 \pi p\over q})\big)\;.\end{align}

Likewise  in the case $n=8$  the  {\em  Mathematica}   output  is
 \begin{align} \nonumber a_{p,q}(8)&=  \frac{1}{6}  q \bigg(1617 + 1512 \cos (\frac{2 \pi  p}{q}) + 462 \cos (\frac{4 \pi  p}{q}) + 
   72 \cos (\frac{6 \pi  p}{q}) + 
   12 \cos (\frac{8 \pi  p}{q}) \\ \label{a8} &+ q\big(-617 - 372 \cos (\frac{2 \pi  p}{q}) - 
    54 \cos (\frac{4 \pi  p}{q})\big)  +q^2 \big(84 + 24 \cos (\frac{2 \pi  p}{q})\big)-4q^3 \bigg)\;.\end{align}
 One has  \begin{align}\label{nicenicenicebis}
a_{p,q}(8)=-\sum_{k_1=1}^q\sum_{k_2=1}^{k_1-2}\sum_{k_3=1}^{k_2-2}\sum_{k_4=1}^{k_3-2}  {{{\tilde{b}}}_{p/q}}(k_1){{{\tilde{b}}}_{p/q}}(k_2){{{\tilde{b}}}_{p/q}}(k_3) {{{\tilde{b}}}_{p/q}}(k_4)\end{align}
%so that $q\ge 7$, if not it trivially vanishes.
  In the reduction of (\ref{nicenicenicebis}) at order $q$ there are  eight  terms with only one sum. They are labelled by  the $2^{8/2-1}=8$ compositions of $4$, $4=4$, $4=3+1$, $4=1+3$, $4=2+2$, $4=1+2+1$, $4=2+1+1$, $4=1+1+2$, $4=1+1+1+1$
\begin{align} \nonumber q[q]a_{p,q}(8)=& +\frac{1}{4} \sum_{k=1}^q {{{\tilde{b}}}_{p/q}}^4(k)\\\nonumber &+ \sum_{k=1}^q {{{\tilde{b}}}_{p/q}}^3(k) {{{\tilde{b}}}_{p/q}}(k-1)\\\nonumber &+ \sum_{k=1}^q {{{\tilde{b}}}_{p/q}}(k) {{{\tilde{b}}}_{p/q}}^3(k-1)\\\nonumber & +\frac{3}{2}  \sum_{k=1}^q {{{\tilde{b}}}_{p/q}}^2(k) {{{\tilde{b}}}_{p/q}}^2(k-1)\\\nonumber &+2  \sum _{k=1}^q {{{\tilde{b}}}_{p/q}}(k) {{{\tilde{b}}}_{p/q}}^2(k-1) {{{\tilde{b}}}_{p/q}}(k-2)\\\nonumber &+ \sum_{k=1}^q {{{\tilde{b}}}_{p/q}}^2(k)
   {{{\tilde{b}}}_{p/q}}(k-1) {{{\tilde{b}}}_{p/q}}(k-2)\\\nonumber & + \sum_{k=1}^q {{{\tilde{b}}}_{p/q}}(k) {{{\tilde{b}}}_{p/q}}(k-1)
   {{{\tilde{b}}}_{p/q}}^2(k-2)\\& + \sum_{k=1}^q {{{\tilde{b}}}_{p/q}}(k) {{{\tilde{b}}}_{p/q}}(k-1) {{{\tilde{b}}}_{p/q}}(k-2) {{{\tilde{b}}}_{p/q}}(k-3)\label{8}\end{align}
   with   output 
\scriptsize\begin{align} \nonumber&{1\over q}\sum_{k=1}^q {{{\tilde{b}}}_{p/q}}^4(k)=2 \big(35 \big)\\\nonumber &{1\over q} \sum_{k=1}^q {{{\tilde{b}}}_{p/q}}^3(k) {{{\tilde{b}}}_{p/q}}(k-1)=2\big(20 + 15 \cos({2 \pi p\over q})\big)\\\nonumber &{1\over q}\sum_{k=1}^q {{{\tilde{b}}}_{p/q}}(k) {{{\tilde{b}}}_{p/q}}^3(k-1)=2\big(20 + 15 \cos({2 \pi p\over q})\big)\\\nonumber & {1\over q}\sum_{k=1}^q {{{\tilde{b}}}_{p/q}}^2(k) {{{\tilde{b}}}_{p/q}}^2(k-1)=2\big(18 + 16 \cos({2 \pi p\over q}) + \cos({4 \pi p\over q})\big)\\\nonumber &{1\over q}\sum _{k=1}^q {{{\tilde{b}}}_{p/q}}(k) {{{\tilde{b}}}_{p/q}}^2(k-1) {{{\tilde{b}}}_{p/q}}(k-2)=2\big(13 + 16 \cos({2 \pi p\over q}) + 6 \cos({4 \pi p\over q})\big)\\\nonumber &{1\over q}\sum_{k=1}^q {{{\tilde{b}}}_{p/q}}^2(k)
   {{{\tilde{b}}}_{p/q}}(k-1) {{{\tilde{b}}}_{p/q}}(k-2)=2\big(12 + 14 \cos({2 \pi p\over q}) + 8 \cos({4 \pi p\over q}) + \cos({6 \pi p\over q})\big)\\ & {1\over q}\sum_{k=1}^q {{{\tilde{b}}}_{p/q}}(k) {{{\tilde{b}}}_{p/q}}(k-1)
   {{{\tilde{b}}}_{p/q}}^2(k-2)=2\big(12 + 14 \cos({2 \pi p\over q}) + 8 \cos({4 \pi p\over q}) + \cos({6 \pi p\over q})\big)\nonumber\\& {1\over q}\sum_{k=1}^q {{{\tilde{b}}}_{p/q}}(k) {{{\tilde{b}}}_{p/q}}(k-1) {{{\tilde{b}}}_{p/q}}(k-2) {{{\tilde{b}}}_{p/q}}(k-3)=2\big(9 + 12 \cos({2 \pi p\over q}) + 9 \cos({4 \pi p\over q}) + 4 \cos({6\pi p\over q}) + \cos({8 \pi p\over q})\big)\;.\label{8bis}\end{align}\normalsize
\subsection*{On the extrapolation}
\noindent  
One can  obtain a  closed  expression for the extrapolated  $a_{p,q}(2j)$'s when   $q<2j$  in terms of the $a_{p,q}(2j)$  defined in (\ref{thea}) when $q\ge 2j$ 
\begin{itemize}
\item
when $ j+1\le q\le 2j-1$: \be a_{p,q}(2j)=0\label{vanish}\ee
\item
when $1\le q\le j$:
\begin{align}
a_{p,q}(2j) =&  \sum_{\substack{k \geq 0}} \sum_{\substack{\ell_1,\ell_2,\ldots,\ell_{\lfloor q/2 \rfloor} \geq 0 \\ \ell_1 + 2\ell_2 + \cdots + \lfloor q/2 \rfloor \ell_{\lfloor q/2 \rfloor} = j-q(k+1)}}
{a_{1,1}\left(2(k+1)\right) \binom{\ell_1+\ell_2 + \cdots + \ell_{\lfloor q/2 \rfloor} + 2k}{\ell_1,\ell_2,\ldots,\ell_{\lfloor q/2 \rfloor},2k}} \prod_{i = 1}^{\lfloor q/2 \rfloor} a_{p,q}(2i)^{\ell_i}\label{soap}
\end{align}\end{itemize}
with \be
a_{1,1}(2j)=  - \sum_{\substack{\ell_1,\ell_2,\ldots,\ell_{j} \geq 0 \\ \ell_1 + 2\ell_2 + \cdots + j \ell_{j} = j}}
\prod_{i = 1}^{j}\frac{ 1}{\ell_{i}!} \left( -{{2i\choose i}^2\over 2i}\right)^{\ell_i}\nonumber\ee
In the RHS of (\ref{soap})  the Kreft coefficient $a_{p,q}(2i)$ is the original one defined for $q\ge 2i$ in (\ref{thea})   whereas in the LHS of (\ref{soap}) the extrapolated Kreft coefficient  $a_{p,q}(2j)$   is computed for $1\le q\le j$.

\noindent The two items above altogether with (\ref{thea}) coalesce to
\begin{equation}\nonumber
a_{p,q}(2j) =  \sum_{\substack{k \geq 0\\j-qk\ge 0}} a_{1,1}\big(2k\big)[z^{2j-q2k}]{{b}}_{p,q}(z)^{1-2k}
\end{equation}
where $ b_{p,q}(z)=-\sum_{j=0}^{\lfloor\frac{q}{2}\rfloor} a_{p,q}(2j)z^{2j}$  is the  Kreft polynomial.  

One also note that the trace formula (\ref{sumformula}) reduces to a partition like formula 
 \begin{align}\label{partition}
\frac{q}{n}\operatorname{Tr} H_{2\pi p/q}^n 
&=  \sum_{\substack{\ell_1,\ell_2,\ldots,\ell_{n/2} \geq 0 \\l_1+2l_2+3l_3+\cdots+(n/2)l_{n/2} = n/2}} \frac{\binom{l_1+l_2+\cdots+l_{n/2}}{l_1,l_2,l_3,\ldots,l_{n/2}}}{l_1+l_2+\cdots+l_{n/2}} \prod_{j = 1}^{ n/2} a_{p,q}(2j)^{\ell_j}
%a_{p,q}(2)^{l_1}a_{p,q}(4)^{l_2} \cdots a_{p,q}({n})^{l_{n/2}}
\end{align} 
when expressed in terms of the extrapolated Kreft coefficients.

\subsection*{Additional binomial identities}
 The binomial countings  (\ref{1000}-\ref{1002}) and (\ref{muiz}) \iffalse, \ref{1003}\fi are particular cases of

      \begin{align} &{l_1+l_2\choose l'_1+l'_2}=\sum_{A=-\infty}^{\infty}{{l_1}\choose {l'_1}+A}{ {l_2} \choose {l'_2}-A }\label{use}\end{align}
    {  the Chu-Vandermonde identity,}
    \begin{align*}&{l_1+l_2+l_3\choose l'_1+l'_2+l'_3}=\sum_{A= -\infty}^{ 
   \infty}\sum_{k_3=-\infty}^{\infty}{ l_1\choose l'_1 - k_3} { l_2\choose
      l'_2 + 2 k_3 + A} { l_3\choose l'_3 - k_3 - A}\;,\end{align*}

 \begin{align*}&{l_1+l_2+l_3+l_4\choose l'_1+l'_2+l'_3+l'_4}=\nonumber\\&
   \sum_{A=-\infty}^{\infty}\sum_{k_3= -\infty}^{ 
   \infty}\sum_{k_4= -\infty}^{ \infty}  {l_1\choose l'_1 - k_4 -   k_3} { l_2\choose 
     l'_2 + k_4 + 2 k_3} { l_3\choose l'_3 +k_4-k_3 + A} { l_4\choose 
     l'_4 - k_4  - A}\;,\end{align*}
 \iffalse 
  \scriptsize \begin{align*}&{2(l_1+l_2+l_3+l_4+l_5)\choose l'_1+l'_2+l'_3+l'_4+l'_5}=\nonumber\\&\sum_{A=-\infty}^{\infty}\sum_{k_3= -\infty}^{ 
   \infty}\sum_{k_4= -\infty}^{ \infty} \sum_{k_5= -\infty}^{ \infty} {2 l_1\choose l'_1 - k_5-   k_4-k_3 } {2 l_2\choose l'_2 -k_5+k_4+2k_3} {2 l_3\choose 
     l'_3 + 4k_5+k_4-k_3} {2 l_4\choose l'_4 - k_5 -k_4 + A} {2 l_5\choose 
     l'_5 - k_5-A } \;,  \end{align*}\normalsize
    \fi and in general of
     \begin{align*} &{l_1+l_2+\ldots+l_j\choose l'_1+l'_2+\ldots+l'_j}=\nonumber\\&\sum_{A=-\infty}^{\infty}\sum_{k_3= -\infty}^{ 
   \infty}\sum_{k_4= -\infty}^{ \infty}\ldots \sum_{k_j= -\infty}^{ \infty} \prod_{i=1}^j{l_i\choose l'_i-k_{i,j} +A(\delta_{i,j-1}-\delta_{i,j})}\;. \end{align*}

\subsection*{Additional  identities}
\noindent One has  also found
\begin{align*}& {1\over q}\sum _{k=1}^q {\tilde{b}_{p/q}}^{{l_1}}(k){\tilde{b}_{p/q}}^{{l_2}}(k-1){\tilde{b}_{p/q}}^{{l_3}}(k-2)+ {1\over q}\sum _{k=1}^q {\tilde{b}_{p/q}}^{{l_2}}(k-1){1\over q}\sum _{k=1}^q {\tilde{b}_{p/q}}^{{l_1}}(k){\tilde{b}_{p/q}}^{{l_3}}(k-2)=\\&2\sum_{k_3=0}^{\infty}{2l_1\choose l_1-k_3}{2l_2\choose l_2+2k_3}{2l_3\choose l_3-k_3}\\\nonumber+&2\sum_{A=1}^{\infty}\cos\left({2A\pi p\over q}\right)
 \sum _{k_3=-[{A/ 2}]}^{\infty } {2
   l_1\choose l_1-k_3-A}{2 l_2\choose l_2+2 k_3+A}
   {2 l_3\choose l_3-k_3}+\\&2\sum_{A=1}^{\infty}\cos\left({2A\pi p\over q}\right)\sum _{k_3=-[{A/ 2}]}^{\infty }{2l_3\choose l_3-k_3-A}
   {2 l_2\choose l_2+2 k_3+A}{2 l_1\choose l_1-k_3} \end{align*}
   and the  binomial identity
 \begin{align*}& {2(l_1+l_2+l_3)\choose l_1+l_2+l_3}+{2l_2\choose l_2}{2(l_1+l_3)\choose l_1+l_3}=\\&2\sum_{k_3=0}^{\infty}{2l_1\choose l_1-k_3}{2l_2\choose l_2+2k_3}{2l_3\choose l_3-k_3}\\\nonumber+&2\sum_{A=1}^{\infty}
 \sum _{k_3=-[{A/ 2}]}^{\infty } {
   2l_1\choose l_1-k_3-A}{ 2l_2\choose l_2+2 k_3+A}
   { 2l_3\choose l_3-k_3}+\\&2\sum_{A=1}^{\infty}\sum _{k_3=-[{A/ 2}]}^{\infty }{2l_3\choose l_3-k_3-A}
   { 2l_2\choose l_2+2 k_3+A}{ 2l_1\choose l_1-k_3} \end{align*}  
 One   derives from (\ref{llbisbis}) %and (\ref{lllbisbis}) 
\begin{align*} \sum_{k=1}^q {\tilde{b}_{p/q}}^{{l_1}}(k){\tilde{b}_{p/q}}^{{l_2}}(k-r)=\sum _{k=1}^q {\tilde{b}_{p/q}}^{{l_1}}(k){\tilde{b}_{p/q}}^{{l_2}}(k-1)\;{\rm with}\; \cos\left({2A\pi p\over q}\right)\rightarrow \cos\left({2rA\pi p\over q}\right)\end{align*}

% \begin{align*} \sum _{k=1}^q {\tilde{b}_{p/q}}^{{l_1}}(k){\tilde{b}_{p/q}}^{{l_2}}(k-r){\tilde{b}_{p/q}}^{{l_3}}(k-2r)=\sum _{k=1}^q {\tilde{b}_{p/q}}^{{l_1}}(k){\tilde{b}_{p/q}}^{{l_2}}(k-1){\tilde{b}_{p/q}}^{{l_3}}(k-2)\;{\rm with}\; \cos\left({2A\pi p\over q}\right)\rightarrow \cos\left({2rA\pi p\over q}\right)\end{align*}
\noindent and  in general  from (\ref{matrix}) 
\begin{align*} &\sum _{k=1}^q {\tilde{b}_{p/q}}^{{l_1}}(k){\tilde{b}_{p/q}}^{{l_2}}(k-r){\tilde{b}_{p/q}}^{{l_3}}(k-2r)\ldots {\tilde{b}_{p/q}}^{{l_j}}(k-(j-1)r)=\\&\sum _{k=1}^q {\tilde{b}_{p/q}}^{{l_1}}(k){\tilde{b}_{p/q}}^{{l_2}}(k-1){\tilde{b}_{p/q}}^{{l_3}}(k-2)\ldots {\tilde{b}_{p/q}}^{{l_j}}(k-(j-1))\;{\rm with}\; \cos\left({2A\pi p\over q}\right)\rightarrow \cos\left({2rA\pi p\over q}\right)\;.\end{align*}
\subsection*{Asymmetric probabilities}
{ Considering   the generalized Kreft coefficient $a_{p,q}^{\lambda}(n)$ obtained in \cite{bis} where  $\lambda/2$ can be interpreted as the ratio  between  the lattice walk probabilities on the horizontal axis versus the vertical axis  and following the same logic as in  (\ref{simplify})  one  writes
\begin{equation}  [q] a_{p,q}^{\lambda}(n)=\frac{1}{n} \sum_A C_n^{\lambda}(A) e^{2 i \pi A p/q}\nonumber\end{equation} to get%\footnote{\bf Or following (\ref{trans}) \begin{align}\nonumber C_n^{\lambda}(A)=&\; n \sum_{l_1, l_2, \ldots, l_{n/2}\;{\rm composition}\;{\rm of}\; n/2} \frac{{l_1+l_2\choose l_1}}{l_1+l_2}\;\; l_2\frac{{l_2+l_3\choose l_2}}{l_2+l_3}\;\ldots \;\; l_{{n/2}-1}\frac{{l_{{n/2}-1}+l_{n/2}\choose l_{{n/2}-1}}}{l_{{n/2}-1}+l_{n/2}}\nonumber\\& \sum_{k_3= 0}^{2l_3 }\sum_{k_4= 0}^{ 2l_4}\ldots \sum_{k_{n/2}= 0}^{2l_{n/2}}\sum_{j_1=0}^{l_1}{l_1\choose j_1}{l_1\choose l_1 +A+\sum_{i=3}^{n/2}(i-2)(k_i-l_i)-j_1}({\lambda\over 2})^{2j_1}\nonumber\\&\sum_{j_2=0}^{l_2}{l_2\choose j_2}{l_2\choose l_2 -A-\sum_{i=3}^{n/2}(i-1)(k_i-l_i)-j_2}({\lambda\over 2})^{2j_2}\prod_{i=3}^{n/2}\sum_{j_i=0}^{l_i}{l_i\choose j_i}{l_i\choose k_i-j_i}({\lambda\over 2})^{2j_i}\;.\end{align}  }
\begin{align}\label{thisistheend} C_n^{\lambda}(A)=&\; n \sum_{l_1, l_2, \ldots, l_{n/2}\atop{\rm composition}\;{\rm of}\; n/2} \frac{{l_1+l_2\choose l_1}}{l_1+l_2}\;\; l_2\frac{{l_2+l_3\choose l_2}}{l_2+l_3}\;\ldots \;\; l_{{n/2}-1}\frac{{l_{{n/2}-1}+l_{n/2}\choose l_{{n/2}-1}}}{l_{{n/2}-1}+l_{n/2}}\nonumber\\& \sum_{k_3= -\infty}^{ 
   \infty}\sum_{k_4= -\infty}^{ \infty}\ldots \sum_{k_{n/2}= -\infty}^{\infty}\prod_{i=1}^{n/2}\sum_{j_i=0}^{l_i}{l_i\choose j_i}{l_i\choose l_i-k_{i,n/2} +A(\delta_{i,n/2-1}-\delta_{i,n/2}
   )-j_i}({\lambda\over 2})^{2j_i}\;.\end{align}  
 $\lambda/2$ is  a deformation parameter  which encapsulates the relative weight of random steps on the horizontal versus the vertical axis. In (\ref{thisistheend})  the $\lambda$-deformation  has narrowed down to replace\footnote{ The same type of  rewriting of (\ref{theformula}) as (\ref{trans}) can be  used for (\ref{thisistheend}).} any binomial ${2l_i\choose l_i'}$ in (\ref{theformula})  by $\sum_{j_i=0}^{l_i}{l_i\choose j_i}{l_i\choose l_i'-j_i}({\lambda/2})^{2j_i}$ (which reduces to ${2l_i\choose l_i'}$ when  $\lambda=2$). Expanding   $C_n^{\lambda}(A)$ in powers of $\lambda$     
\be \nonumber C_n^{\lambda}(A)=\sum_{m=0}^{n/2} C_{m,m,n/2-m,n/2-m}(A)({\lambda\over 2})^{2m}\ee  
yields $C_{m,m,n/2-m,n/2-m}(A) $, the number of  closed lattice  walks of length $n$ with $m$ steps  right, $m$ steps left, $n/2-m$ steps up, $n/2-m$ steps  down  enclosing a given algebraic area $A$. 
It is obtained by extracting  the coefficient  at order  $({\lambda/ 2})^{2m}$ in  (\ref{thisistheend}).
}

\end{document}